\documentstyle[psfig]{cupconf}

\ifoldfss
\else
  \ifnfssone
    \newmathalphabet{\mathit}
      \addtoversion{normal}{\mathit}{cmr}{m}{it}
      \addtoversion{bold}{\mathit}{cmr}{bx}{it}
    \newmathalphabet{\mathcal}
      \addtoversion{normal}{\mathcal}{cmsy}{m}{n}
    \else
    \ifnfsstwo
    \fi
  \fi
\fi

\def\hexnumber#1{\ifcase#1 0\or1\or2\or3\or4\or5\or6\or7\or8\or9\or
 A\or B\or C\or D\or E\or F\fi }

\makeatletter
\ifx\CUP@mtlplain@loaded\undefined
\else
\fi
\makeatother
\makeatletter
\ifx\CUP@mtlplain@loaded\undefined
\else
\fi
\makeatother
\title[$H_2$ molecules and cold clouds]{$H_2$ molecules and cold clouds
in cooling flow clusters\footnote{\hrule \vskip2mm
Poster presented at the International 
Conference {\it $H_2$ in space}, IAP-Paris (France)
\\ September 18$^{th}$-October 1$^{st}$, 1999}
}

\author[L. Grenacher, Ph. Jetzer, D. Puy]
{L. Grenacher, Ph. Jetzer, D. Puy}

\affiliation{$^1$Paul Scherrer Institute, Laboratory for Astrophysics, 
Villigen (Switzerland) \\[\affilskip]
$^2$Institute of Theoretical Physics, University of Zurich (Switzerland)}

\begin{document}
\ifnfssone
\else
  \ifnfsstwo
  \else
    \ifoldfss
      \let\mathcal\cal
      \let\mathrm\rm
      \let\mathsf\sf
    \fi
  \fi
\fi

\maketitle

\begin{abstract}

\end{abstract}

\firstsection 
\section{Introduction}

Around the epoch of recombination atomic hydrogen is the most 
important chemical species and leads by adiabatic cooling of the universe
to the formation of molecular hydrogen $H_2$ (Puy et al. 1993, 
Puy \& Signore 1999). The actual 
$H_2$ content
is very uncertain and estimated only indirectly. The important recent observation of
the lowest pure rotational lines of $H_2$ in the spiral galaxy NGC 891
(Valentijn \& Van der Werf 1999) gives a direct 
indication of relatively 
warm (T=150-230 K) molecular clouds in the disk in addition to a massive
cooler (80-90 K) component in the outer regions.
\\
In clusters of galaxies X-ray measurements show an excess absorption below 
$\sim$ 1 keV compared to a best fit bremsstrahlung model, which is 
interpreted as due to the presence of cold molecular clouds (White et al. 
1991).
\\
In a scenario with successive fragmentation of these clouds we calculate
the molecular rotational line cooling due to $HD$- and $H_2$-molecules
and determine their minimum temperature achievable in equilibrium with
the exterior bremsstrahlung of the hot intracluster gas.

\section{Molecular cooling} 
The abundance $\eta_{HD}$  of the $HD$ molecule is considered to be 
primordial:
$$
\eta_{HD}\, \simeq \, 7 \times 10^{-5}, \ {\rm Signore \, \& \, Puy \ 1999.}
$$
The quadrupole transitions of $H_2$ are
supposed to follow an ortho/para ratio of 1, its  first excited state is 
at 512 K. Although this cooling is less efficient than the cooling due to 
$HD$, $H_2$ is more abundant and we expect the $H_2$ cooling to be 
more important than the $HD$ cooling in the temperature region above $\sim$
100 K and in the density region of $10^4$ cm$^{-3}$.
\\
Considering only the transition between the ground state and the first 
rotational level, the molecular cooling can be calculated analytically 
(Puy, Grenacher and Jetzer 1999). In the case of $H_2$-clouds temperatures
as mentioned above, the higher excitation levels turn out to be significant
even if they are still very weakly populated.
\\
In Figure 1 we show for the total cooling $\Lambda_{H_2}+\Lambda_{HD}$
the relative importance of the cooling agents $H_2$ and $HD$
\begin{eqnarray}
\alpha_{H_2}&=&\frac{\Lambda_{H_2}}{\Lambda_{H_2}+\Lambda_{HD}} \nonumber \\ 
\alpha_{HD}&=&\frac{\Lambda_{HD}}{\Lambda_{H_2}+\Lambda_{HD}}~.\nonumber
\end{eqnarray}
We can see clearly,
that $H_2$ becomes the most important cooling agent above $\sim$ 70 K.

\begin{figure}
\centerline{\psfig{figure=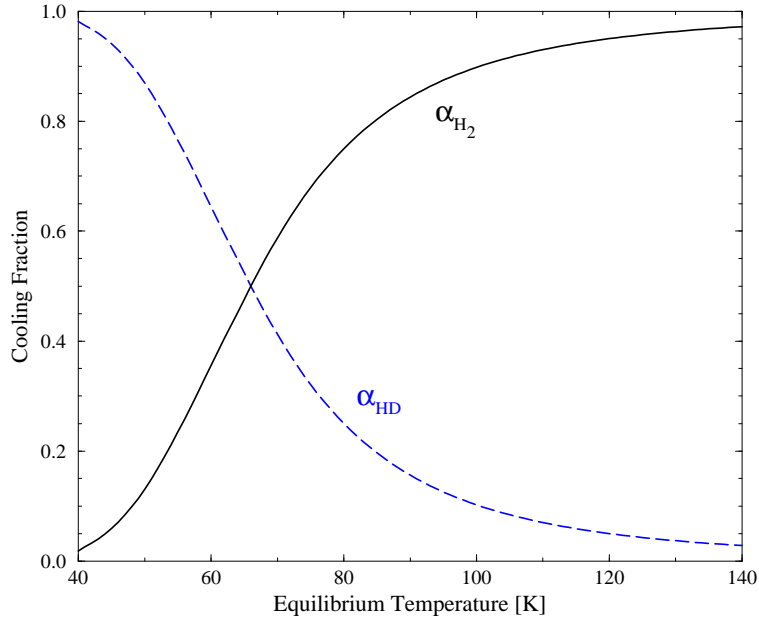,width=10cm,angle=-90}}
    \caption{\small{Relative importance of the different cooling agents $H_2$,
$\alpha_{H_2}=\Lambda_{H_2}/(\Lambda_{H_2}+\Lambda_{HD})$ and $HD$,
$\alpha_{HD}=\Lambda_{HD}/(\Lambda_{H_2}+\Lambda_{HD})$.}}
\end{figure} 
\section{Thermal equilibrium}
The most important heat source for molecular clouds in cooling flows is the 
X-ray bremsstrahlung emitted by the hot intracluster gas.
In the successive fragmentation scenario this radiation is shielded by
the presence of an attenuating column density in the outher parts of the 
clouds, which we take into account by the attenuation 
factor, $\tau$, following O'Dea et al. (1994). 
\\
The balance between heating and 
cooling in the cluster environment leads to a thermal equilibrium inside
the cooling flow region of the clusters. 
The escape probability for this molecular emission is close to 1. We 
investigate the coldest equilibrium achievable inside the cooling flow
region. 
\section{Discussion}
We consider small clouds ($\sim 10$ AU) with an $H_2$ density of 10$^4$ 
cm$^{-3}$, attenuated by $\tau=0.01$. 
The X-ray heating ($\Gamma_X(r)\propto r^{-3})$ is of course more important in 
the cluster center and low equilibrium temperatures are 
achieved at large distances from the cluster center. Nevertheless, 
the fact that the cold clouds are located in the cooling flow region
implies that 
the distance must be below the cooling radius. In this context we have 
calculated the equilibrium temperature of the molecular clouds located at 
the cooling radius. The table gives these equilibrium temperatures $T_{clump}$ 
for different clusters of galaxies.

\begin{table}
   \begin{center}
   \begin{tabular}{lccc}
      {\bf Cluster} & {${\mathbf T_{clump}}$} & {${ \mathbf r_{cool}}$} & {${\mathbf T_{Kev}}$}\\
             & {(in K)} & {(in kpc)} & {(in keV)}\\
              \hline
              Centaurus    & 153 & 87  & 2.1\\
              Hydra A      & 107 & 162 & 4.5\\
              PKS 0745-191 & 92 & 214 & 8.6\\
              Abell 262    & 119 & 67 & 2.5\\
              Abell 426    & 83 & 145 & 6.3\\
              Abell 478    &228& 240 & 7.1\\
              Abell 496    &58& 138 & 4.8\\
              Abell 539    &133& 34 & 3.4\\
              Abell 576    &102& 69 & 2.9\\
              Abell 1060   &102& 68 & 3.3\\
              Abell 1367   &80& 40 & 4.1\\
              Abell 1795   &136& 181 & 5.3\\
              Abell 2052   &96& 140 & 3.4\\
              Abell 2151   &256& 146 & 2.9\\
              Abell 2159   &128& 119 & 4.5\\

    \end{tabular}
    \end{center}
    \caption{The equilibrium temperature $T_{clump}$ at the cooling radius
$r_{cool}$ for the cluster temperature $T_{keV}$ are shown for different
cluster of galaxies.}
\end{table}

In the region of the cooling flow, we find that an equilibrium is 
possible at low temperature (below 200 K) due to $H_2$ cooling.
\\
Whether the cloud can be cooled down to $\sim$ 70 K, where $HD$ dominates, 
depends on various parameters, such as the cluster temperature $T_{keV}$,
the attenuation factor $\tau$, the cooling radius $r_{cool}$ and other
characteristics of the hot intracluster gas.
\\
The detection of these $H_2$ molecules is difficult, because the first 
rotational level, acccessible only through a quadrupolar transition, is 
more than 500 K above the fundamental. The study of CO-$H_2$ ratios can
give some insight, because the $CO$ molecules are excited by 
collisions with $H_2$, and should be a tracer of cold $H_2$ clouds 
(Grenacher et al. 1999). In this 
context cold $H_2$-clouds could be an interesting possibility of baryonic 
dark matter (Combes 1999). The FUSE satellite will certainly give clarification on this 
problem.

\begin{acknowledgments}
We are grateful to acknowledge Francoise Combes for organizing such a pleasant 
conference.  This work has been supported by the {\it D$^r$ Tomalla 
Foundation} and by the Swiss National Science Foundation.
\end{acknowledgments}


\begin{thebibliography}{} 

\bibitem[Combes (1999)]{combes}
     {\sc Combes , F.} 1999
      {\it astro-ph/9910296} SISSA-Babbage
\bibitem[Grenacher et al. (1999)]{gre}
     {\sc Grenacher, L., Jetzer Ph. \& Puy, D.} 1999
      {\it work in progress}              
\bibitem[O'Dea et al. (1994)]{ode}
     {\sc O'Dea, C., Baum, S., Maloney, P. et al.} 1994
     {\it ApJ} {\bf 422}, 467
\bibitem[Puy et al. (1993)]{puy1}
     {\sc Puy, D., Alecian, G., Lebourlot, J. et al.} 1993
     {\it A\&A} {\bf 267}, 337 
\bibitem[Puy \& Signore (1999)]{puy3}
     {\sc Puy, D. \& Signore, M.} 1999
     {\it New Astr. Rev.} {\bf 43}, 223 
\bibitem[Puy et al. (1999)]{puy4}
     {\sc Puy, D., Grenacher, L. \& Jetzer, Ph.} 1999
     {\it A\&A} {\bf 345}, 723
\bibitem[Signore \& Puy (1999)]{signore}
     {\sc Signore, M. \& Puy, D.} 1999
     {\it New Astr. Rev.} {\bf 43}, 185 
\bibitem[Valentijn \& Van der Werf (1999)]{val}
     {\sc Valentijn, E. \& Van der Werf, P.} 1999
      {\it ApJ} {\bf 522}, L29
\bibitem[White et al. (1991)]{white}
     {\sc White D., Fabian A., Johnstone R. et al.} 1991
     {\it MNRAS} {\bf 252}, 72   
\end{thebibliography}
\end{document}